\documentclass[prd,tightenlines,floatfix,preprint,nofootinbib,eqsecnum,12pt]{revtex4-1}

\begin{document}

\title{The Xerographic Distribution: \\ Scientific Reasoning in a Large Universe\footnote{
Invited talk by M. Srednicki at {\it Quantum Theory and Symmetries 6}, Lexington, KY,
20-25 July 2009.}}

\author{Mark Srednicki}
\email{mark@physics.ucsb.edu}

\author{James Hartle}
\email{hartle@physics.ucsb.edu}

\affiliation{Department of Physics, University of California, Santa Barbara, CA 93106}

\date{\today}

\begin{abstract}
As observers of the universe we are physical systems within it. If the universe is very large in space and/or time, the probability becomes significant that the data on which we base predictions is replicated at other locations in spacetime.  Predictions of our future observations therefore require an assumed probability distribution---the {\it xerographic distribution}---for our location among the possible ones. It is the combination of basic theory plus the xerographic distribution that can be predictive and testable by further observations.  This is illustrated
by examining a toy model of a classical deterministic universe with a fixed flat metric.
\end{abstract}


\maketitle

Many contemporary cosmological models predict that the universe is much larger than our observable Hubble volume, and is possibly infinite.  In such a large (or infinite) universe,
the probability that there are multiple copies of `our' data, $D_0$, approaches (or equals) 
100\%.  

The data $D_0$ consists of every scrap of information that humanity possesses, including all memories and all records of every observation, individual and collective, that has ever been made about anything. `Humanity' refers to a particular physical subsystem of the universe, consisting of human beings and all associated technology. Terms such as `our', `we', etc., refer to this subsystem.

The possible existence of multiple copies of $D_0$ presents a challenge to the conventional use of scientific reasoning \cite{SH09}. 
To see why, consider a hypothetical situation in which our data $D_0$ happens to be consistent with a completely classical dynamical theory (of particles 
and/or fields), and a fixed, flat spacetime metric.   In such a situation, a complete physical theory $T$ consists of the dynamical equations (for the particles and/or fields) and an initial condition.  

A theory $T$ is compatible with our data $D_0$ if it predicts the occurrence of {\it at least one\/} instance of $D_0$ throughout spacetime \cite{HS07}.
In order to avoid issues of infinities, we suppose that $D_0$ is compatible with at least some theories $T$ with finite spatial volume and finite temporal extent (in the coordinate system that makes the dynamical equations as simple as possible).  

If a theory $T$ predicts {\it exactly one\/} instance of $D_0$ anywhere in the finite spacetime, then we can use $T$ to make predictions about our future observations, and hence test $T$.

If, however, a theory $T$ predicts {\it more than one\/} instance of $D_0$ (at various locations in spacetime), and furthermore predicts that {\it different sets of future observations\/} will occur at each of these locations, then we cannot use $T$ to make definite predictions about our own future observations. This is because we do not know which copy of $D_0$ we correspond to; we could be any one of them. 

Note that this is not due to any incompleteness of the theory $T$; it makes definite predictions for what will happen at each of the instances of $D_0$.  These predictions can be called
{\it third-person\/} predictions.  Third-person predictions could be completely verified only by a hypothetical `third-person' observer outside the system described by $T$.  We, however, are observers within the system.  Therefore, what we would like to have is a {\it first-person\/} prediction of what `we' will see next.  The theory $T$, though fully complete and deterministic in the usual sense, cannot provide such a prediction by itself.  

To make a prediction of what `we' will see next, we need to decide which copy corresponds to `us', or, more generally, posit a probability distribution---the {\it xerographic distribution\/}---on the set of copies.  More specifically, if the theory $T$ predicts that there are $N$ copies of $D_0$ throughout the finite spacetime, then we assign probability $\xi_i$
that the $i$th copy (in some numbering scheme) corresponds to us.  Given a
{\it theoretical framework\/} $(T,\xi)$ consisting of {\it both} a theory $T$ {\it and\/} a xerographic
distribution $\xi$, probabilistic predictions of what `we' will see next can be made.  

Now the problem is how to choose the xerographic distribution $\xi$.  
The most important point to remember is that 
$\xi$ is {\it not\/} determined by the theory $T$, which predicts the future of each instance of 
$D_0$, but does not tell us which instance is ours.  Therefore, the xerographic distribution
is an {\it additional ingredient\/} that must be posited, along with a conventional theory $T$,
in order to have a framework $(T,\xi)$ that is capable of generating first-person predictions for
(probabilities of) what `we' will see.  

The simplest and most obvious choice of $\xi$ is uniform on the possibilities:
$\xi_i=1/N$.  This can be considered as an application of the `principle of indifference':
we have no reason to choose one copy over another, and therefore we should assign
them all equal probability.

There are, however, situations in which we may indeed have a good reason for choosing
one copy over another.  

Suppose that some of the copies are `Boltzmann brains' (BBs for short) \cite{BBrefs}, 
in which the data $D_0$ is not a faithful record of actual events and circumstances, but simply a chance fluctuation.  Other copies are `ordinary observers' (OOs for short), where the data does faithfully record actual events and circumstances. BBs and OOs can be considered as endpoints of a continuum of some measure of the faithfulness of records and memories to external reality; the degree of faithfulness of a given copy could in principle be determined by a third-person observer.  For simplicity of exposition we will divide all copies of $D_0$ into OOs (which have a high degree of faithfulness) and BBs (which have a low degree of faithfulness).

We may wish to discount the possibility that we are BBs.  This is because we believe that  we currently have a coherent picture of the universe and at least some of the laws that govern it; if we are BBs, this picture is fraudulent.  Therefore, whatever evidence led us to consider the theory $T$ (which predicts the existence of BBs) cannot be trusted, and there is no reason to favor $T$ over any other theory that allows for BBs with our data $D_0$, including theories that have a completely different external reality.  The belief that we are BBs is therefore `self-undermining' \cite{Car04}.  

We can avoid the problem of possibly holding self-undermining beliefs by choosing $\xi_i=0$ for any copy $i$ which is a BB. This choice has the nature of a `working assumption'.  We choose to {\it assume\/} that our memories and records faithfully reflect external reality.  Such an assumption implicitly underlies all of science.  

To see the consequences of these different possible choices for $\xi$, consider a theory $T$ that predicts the existence of one OO consistent with $D_0$, and also a large number of BBs.  Furthermore, while the future of the OO is a straightforward continuation
of its past, the futures of the BBs are all wildly different, and inconsistent with their (faulty) records.  

If we take a uniform $\xi$, then we predict that it is overwhelmingly likely that `our' future is 
very unlike our past.  Let's now wait a few seconds $\ldots\ $ \cite{Got08}; we note that our present has turned out to be quite like our past, and so this framework---the theory $T$ that predicts one OO and lots of BBs, and the uniform xerographic distribution---is ruled out with a high degree of confidence.  

If, on the other hand, we take 
$\xi_{i={\rm\scriptscriptstyle OO}}=1$ and $\xi_{i={\rm\scriptscriptstyle BB}}=0$, then
we predict that `our' future is very much like our past.  
Let's now wait a few seconds $\ldots\ $.
We note that the present is quite like the past, and so this framework---the theory $T$ that predicts one OO and lots of BBs, and the xerographic distribution that is peaked on the 
OO---is fully supported by the subsequent data.

More generally, different choices of $\xi$ can be treated in the same manner as different choices of $T$. Once a framework $(T,\xi)$ is specified, we can see if it makes any high-probability predictions of what `we' will see, and, if so, test these predictions.  
If the predictions are successful,
the framework $(T,\xi)$ remains a viable one.   If the predictions are unsuccessful, a new
framework (with either a different theory $T$ or a different xerographic distribution $\xi$
or both) must be postulated, and the process of prediction and testing repeated.  The details
of a standard Bayesian analysis of this scientific process are given in \cite{SH09}. 

Some scientists believe that the uniform $\xi$ must always be chosen (see e.g.~\cite{GV07}), 
so that if a 
framework $(T,\xi_{\rm uniform})$ makes an incorrect prediction, then the theory $T$ must
be discarded, and $\xi_{\rm uniform}$ retained.  The present authors
see no fundamental justification for this rigid philosophical stance.  {\it Any\/} of our
scientific assumptions might be wrong, and so differing assumptions should be competed 
and compared on various criteria, including how predictive they are.  

Modern cosmological theories must also include the complicating features of 
both quantum mechanics and general relativity.  Since quantum mechanics is inherently
probabilistic, and provides a scheme for computing probabilities, it might seem that the
xerographic distribution is computable from the underlying theory.  As our purely
deterministic example shows, however, this is not possible.  Quantum mechanics is no
more capable than classical mechanics of telling us which copy is `ours', provided that
there is a nonzero third-person probability of the existence of more than one copy.  
General relativity complicates the issue by making the counting of copies dependent
on the choice of a reference volume.  This is especially problematic in the case of an infinite universe; 
some sort of cutoff procedure must be adopted, numbers of copies computed, and then appropriate limits taken.

How to deal with the complications of quantum mechanics and general relativity (in, say,
computing the ratio of the number of OOs to BBs) comprises the `measure problem'
of modern cosmology \cite{measrefs}.  Some scientists hope that further theoretical developments in our understanding of quantum gravity will solve this problem---effectively, 
tell us which reference volumes to use and how to regulate infinities.  
Whether or not this is the case,
we will still be left with the issue of comparing and competing different xerographic
distributions on the (now suitably regulated and well-defined) set of copies of our data.  

Thus, in a universe containing more than one copy of our data,
the set of physical theories under consideration must be supplemented by a set of
xerographic distributions under consideration.  Best scientific practice dictates that
a rigid dogmatic choice of either should be avoided.  

\acknowledgments
We thank Raphael Bousso, Brandon Carter, Sean Carroll, Ben Freivogel, Steve Giddings, 
Alan Guth, 
Thomas Hertog, Matthew Kleban, Don Page, Steve Shenker, and Alex Vilenkin for numerous helpful conversations.
This work was supported in part by the National Science Foundation under grants PHY05-55669
and PHY07-57035.

\end{document}